**Evidence of catalytic production of hot hydrogen in rf generated hydrogen/argon plasmas**

Jonathan Phillips[a]

University of New Mexico, Department of Chemical and Nuclear Engineering,

Farris Engineering Center, Albuquerque, NM 87131

Chun-Ku Chen, Kamran Akhtar, Bala Dhandapani and Randell Mills

BlackLight Power, Incorporated, 493 Old Trenton Road, Cranbury, NJ 08512

This is the third in a series of papers by our team on apparently anomalous Balmer series line broadening in hydrogen containing RF generated, low pressure (< 600 mTorr) plasmas. In this paper the selective broadening of the atomic hydrogen lines in pure $H_2$ and Ar/$H_2$ mixtures in a large 'GEC' cell (36 cm length × 14 cm ID) was mapped as a function of position, $H_2$/Ar ratio, time, power, and pressure. Several observations regarding the selective line broadening were particularly notable as they are unanticipated on the basis of earlier models. First, the anomalous broadening of the Balmer lines was found to exist throughout the plasma, and not just in the region between the electrodes. Second, the broadening was consistently a complex function of the operating parameters particularly gas composition (highest in pure $H_2$), position, power, time and pressure. Clearly not anticipated by earlier models were the findings that under some conditions the highest concentration of 'hot' (>10 eV) hydrogen was found at the entry end, and not in the high field region between the electrodes and that in other conditions, the hottest H was at the (exit) pump (also grounded electrode) end. Third, excitation and electron temperatures were less than one eV in all regions of the plasma not directly adjacent (>1mm) to the electrodes, providing additional evidence that the energy for broadening, contrary to standard models, is not obtained from the field. Fourth, in contrast to our earlier studies of hydrogen/helium and water plasmas, we found that in some conditions 98% of the atomic hydrogen was in the 'hot' state throughout the GEC-type cell. Virtually every operating parameter studied impacted the character of the hot H atom population, and clearly second and third order effects exist, indicating a need for experimental design. Some non-field mechanisms for generating hot hydrogen atoms, specifically those suggested by Mills' CQM model, are outlined.

[a] Corresponding author: phone 505-665-2682; fax 505-665-5548 jphillips@LanL.gov

## I. INTRODUCTION

For more than 15 years there have been reports of selective broadening of the Balmer series α line in low pressure (ca. <1 Torr) pure hydrogen, and RF and DC generated $H_2$/Ar plasmas [1-10]. New reports continue to appear [11], and now there are several reports from our team that selective $H_\alpha$ line broadening is found in other plasmas called resonant-transfer (rt)-plasmas including $H_2$/He (RF-GEC-type cell, [12]), water (microwave [13] and RF-GEC-type cell [14]) and in the presence of certain catalysts including helium, argon, potassium, and strontium (glow discharge, RF-GEC, and filament-type cells [15-22]). In all prior cases, other than those from our team [12–22], the selective broadening was studied exclusively for hydrogen and $H_2$/Ar plasmas and with one exception [23] exclusively in high field regions between the electrodes. With one exception [24] no studies outside our group employed microwave systems.

In all previous reports it is noted the broadening was limited to hydrogen atoms (e,g, no broadening of Ar ion or hydrogen molecular lines), and there is universal agreement regarding the origin of the broadening, Doppler shift. Generally, there is agreement that the energy of the 'hot' hydrogen between the electrodes is greater than 15 eV.

Outside of our team there is almost universal agreement about the mechanism of energy input to the hydrogen, although recent observations have led some of the proponents of the theory to make significant additions. Specifically, apart from very recent changes proposed by one team [23], and our own consistent support for the CQM model there has been universal support in the literature for a 'field acceleration' origin for the hot hydrogen (see Discussion section). It should be noted, however, that there are variations on the details of the mechanism that selectively provide energy to atomic H. Some groups postulate that molecular hydrogen ions absorb energy directly from the field, undergo charge exchange, and then dissociate. Others, in order to explain the symmetry of the broadened lines (inconsistent with acceleration along a single axis) suggest that the hot atoms are formed when hydrogen species adsorbed on the electrode are bombarded by hot accelerated ions, resulting in the formation of hot hydrogen atoms. However, in all these models, the energy to generate the hot hydrogen clearly comes from ions absorbing energy from the field. The unusual nature of the observed phenomenon, and the apparently novel twist to standard plasma physics required by the 'field acceleration' models, inspired the present work.

In particular, it is not clear that the 'field acceleration' class of models can explain some of the phenomenon already observed: (i) a bimodal (or trimodal) distribution of neutral species temperatures, (ii) neutral species temperatures more than thirty times higher than those of electrons in an RF plasma, (iii) symmetric Doppler broadening even in the absence of reflecting surfaces, and iv) recent, but limited, indications that the angle of observation relative to the direction of field has no impact on the characteristic of the broadened line. Nowhere else in the literature of plasma physics is there a mechanism postulated that creates a body of neutral species hotter than any of the charged species. Certainly nowhere else is there a mechanism postulated for generating neutral species that are far hotter than electrons in any type of plasma. Moreover, recent experimental results in the literature suggests symmetric line broadening even in the absence of reflecting surfaces appears to contradict the predictions of the model and appear to have inspired some dramatic changes proposed for the model by one team [23].

Given recent data suggesting broadening in low field regions and rapid rate of proposed changes to the models, it is clear there is a need for additional data to further test the entire class of 'field acceleration' models. In this study the **scientific method** was employed to test clear predictions of the field acceleration models. It must be noted that the scientific method requires that given the failure of the dominant paradigm to explain data, consideration be given to alternative models. Specifically, we argue that an alternative model already exists, that it is entirely consistent with all observations of selective line broadening, (as well as many other phenomenon) and hence must be considered.

The method of testing the field acceleration class of models required an experimental design both more thorough, and broader than prior studies of Balmer series line broadening in low pressure (between 100 and 600 mTorr) RF generated $H_2$/Ar plasmas. For example, the field acceleration models all require a high field to produce (selectively) high energy atomic hydrogen. Thus, the experiments conducted for this study were designed to search for Balmer series line broadening clearly far outside the high field regions and also beyond the thermalization distance from any possible fast-H-forming regions possibly due to high electric fields. Also, the field acceleration model predicts that the observed line broadening should be impacted by the angle of orientation relative to the accelerating field. Observation perpendicular to the field should show little broadening and that parallel to the field should show a preferential red or blue shift as a function of alignment relative to the cathode/anode. Thus, the experiment was designed to permit observation parallel and perpendicular to the principle field directions.

In brief, all the results of these particular tests of the field acceleration models were contrary to the expectations of the field acceleration class of models. Line broadening of undiminished magnitude was found throughout the plasma, both in regions of high field and regions of low field. In some cases the intensity of the broadened Balmer lines was in fact greatest in regions of low field. Also, absolutely no effect of orientation relative to field direction on the symmetry of the Doppler broadened line was found.

In contrast, all the observations are consistent with the Mills' classical quantum mechanics (CQM) model. That is, CQM is consistent with (i) hot hydrogen forming everywhere, (ii) the finding that the hot H concentration profile is not related to the field strength within the system, (iii) the finding that the neutral H atoms are very much hotter than the electrons, (iv) the finding that the Doppler energy of the hot atomic hydrogen is a function of position, time, gas composition and ratio, (v) in fact the hottest H atoms are in some cases found at the grounded electrode end of the plasma, (vi) the increased population of H atoms in particular plasmas and (vi) the failure to find line broadening in $Xe/H_2$ plasmas. Moreover, regarding anomalous Doppler broadening of atomic hydrogen emission lines, consistency of observations with the CQM model, and not with the accelerating fields class of models, is not unique to the present results. Indeed, earlier reports from our group regarding anomalous Doppler broadening of atomic hydrogen emissions in systems with virtually no field [17-22], in microwave plasmas [13, 15-16], as well as in RF-GEC systems with $He/H_2$ [12] and $H_2O$ [14] plasmas. In none of these examples are the results of the spectroscopy at all consistent with predictions of the accelerating fields class of models. For example, those models cannot be applied to microwave plasmas as it is well known that ions do not pick up energy from the field in a microwave system.

## II. EXPERIMENTAL

### A. Plasma Hardware

As described in earlier reports, all plasmas were generated in a GEC-type cell [15-16] held between 100 and 600 mTorr. This system consists of a large cylindrical (14 cm ID × 36 cm length) Pyrex chamber with two steel circular (8.25 cm diameter) plates, placed about 1 cm apart at the center (Figure 1). A radio frequency generator at 13.6 MHz (RF VII, Model RF 5) was used to couple power to the plates using an efficient impedance matching network. A stainless steel tube (ID 0.95 cm) supplied power to the hollow powered electrode. This arrangement facilitated the end-on observation of the plasma between the electrodes in the direction parallel to the electric field. All parts, chamber, power supply, gauges, spectrometer etc. were grounded with heavy-duty aluminum foil, as this was found to dramatically improve signal to noise ratio. Gases, UHP grade (99.999%) Ar, $H_2$ and Xe, were metered into the chamber through Ultratorr fittings at one end, about 18 cm from the electrodes, using independent mass flow controllers (MKS) for each gas. To test for the purity of the $H_2$ only plasmas, efforts were made to find the Ar line at 696 nm. Absolutely no argon signal was detected in the pure hydrogen plasmas.

The chamber was pumped using a Welch two-stage rotary vane oil sealed vacuum pump (Model 8920) with a rated capacity of 218 L/min. This pump was attached to the chamber with a 1 cm ID Ultratorr fitting at the end opposite that at which gas entered. Pressure was measured with an MKS Baratron gauge placed between the pump and the plasma chamber. In order to maintain the pressure levels desired (200-600 mTorr) it was necessary to keep the total flow rate low, between 0.2 and 2 $cm^3$/min STP in all cases. The equivalent 'volume' of gas in the GEC-type cell, corrected for pressure (estimated 500 mTorr) and temperature (estimated 423 K), is about 0.35 $cm^3$ STP.

The RF system is designed for the control of forward (absorbed) power. In all cases that level was set to between 100 and 200 Watts, as noted. Curves of measured voltages as a function of power and gas composition are given in Figure 2.

**B.     Spectrometer**

The spectrometer system, a 1.25 m visible light instrument from Jvon-Spex with a holographic ruled diffraction grating (1800 g mm$^{-1}$), that has a nearly flat response between 300 and 900 nm, is described in detail elsewhere [15]. The slit was set at 10 μm in all cases and light was collected using a light fiber bundle consisting of nineteen 200 μm fibers. This fiber was placed either near an end of the chamber, approximately 15 cm from either electrode (Positions 1 and 3, Figure 1), in a quartz insert tube 1 cm in diameter, that protruded 2 cm into the cell or in a few cases in hollow grounded electrode with a quartz window, such that the probe collected light on a line parallel to the field direction at the center of the space between the electrodes (Position 2-P).

The orientation of the probe relative to the azimuthal axis was carefully controlled. In Position 2, the fiber was pointed directly at the centerline. In Position 2b it was oriented directly along that axis. In contrast, at Positions 1 and 3 the probe was placed just above the glass of the cell vertically, and about 7 cm, directly above the electrode power feeds. It was then 'aimed' by selecting the angle relative to the line joining the fiber probe and the electrode feed. In most cases, the angle was chosen to be 30°, although a study of the impact of this 'tilt angle' on the measured broadening was conducted (more later). In order to demonstrate that orientation relative to the azimuthal axis of the chamber is significant in controlling the origin of light that is collected, tests with a red laser were performed. These clearly showed, consistent with the listed 9° acceptance angle specification, that the probe (placed on 'top') really only sees light originating immediately below it from a cone with a maximum approximate diameter of 1 cm at the opposite end ('bottom') of the GEC type cell. Thus, the light originated 15 ± 1 cm from the electrodes at Points 1 and 3.

Broadened lines were fit using two or three Gaussians, one for the fast ('hot'), one for the 'warm' (when detected) and the third for slow ('cold') hydrogen, using a procedure described elsewhere [12]. It is notable that the curve fittings achieved were excellent ($R^2 > 0.97$ in all cases).

**C. Electron Energy**

Electron energy was determined using a standard emission spectroscopy method [25-27]. The method requires measuring the line intensity of an isolated line and the adjacent continuum intensity. For this work, the line selected was the argon 430 nm. This line is generally chosen because there is no interference from other lines, the transition probability (Einstein A factor) is well known, and other factors (see Table IV) are independent of temperature at this wavelength. In short, only the relative intensity of the line and the continuum must be measured.

**III. RESULTS**

In evaluating the veracity of the work performed it is important to note that in areas of experimental overlap, there was no significant difference between observations made in the present work and those made in earlier studies of line broadening [1-11,23-24]. In agreement with earlier studies of $H_2$/Ar plasmas it was found that only the Balmer series ($\alpha$, $\beta$, $\gamma$) lines are broadened. Consistent with earlier reports, all data indicate the broadening must be Doppler. And, like earlier work, there is no indication that the angle of observation relative to the field direction plays any role in the outcome.

There are twelve significant quantifiable results, and except as noted, they are unreported in prior studies by other groups: One: Consistent with earlier observations, line broadening is limited to the atomic hydrogen lines. Two: Hot hydrogen of Doppler energy greater than 10 eV is found throughout the GEC-type cell, and not only between the electrodes. Three: Several operating parameters influence the magnitude of the broadening. For example, the highest H atom temperatures (60 eV) were found in pure hydrogen at low pressure (80 mTorr) run for long-duration. At one set of conditions, the magnitude of the broadening was studied extensively as a function of operating conditions and was found to reach a maximum value as a function of the amount of Ar present. Four: The temperature of hot hydrogen under many operating conditions is hottest at Position 3, the pump end of the GEC-type cell. Five: Under some circumstances the concentration of hot hydrogen is highest at the gas entry point, and decreases toward the gas exit point. In contrast, the concentration of cold hydrogen under all circumstances is highest between the electrodes, and decreases symmetrically toward both ends. Six: Consistent with a limited volume of earlier data, there is no impact of orientation relative to field direction on the symmetry of the broadened lines. Seven: It was shown again that in control mixtures ($H_2$/Xe) there is no broadening. Eight: It is demonstrated that the fraction of atomic hydrogen in the 'hot' state increases with time under some conditions. Nine: The fraction of hot H increases with Ar content and in 'high' Ar fraction plasmas it can reach nearly 99%. Ten: The maximum energy value of hot hydrogen in pure $H_2$ plasmas increases with decreasing pressure and increasing time. Eleven: The temperature of the electrons, consistent with earlier reports, is less than 1 eV throughout the cell. Twelve: The measured electron temperature, and the measured hydrogen excitation temperature are very nearly the same, as expected for an RF plasma.

**A. Line Broadening**

Consistently the Balmer lines in both the Ar/H$_2$ and H$_2$ only plasmas have two or three components: cold (<0.2 eV), warm (<3 eV) and hot (>10 eV). As shown in Tables I–III, the early (<10 hours on stream) plasmas, at least for the gas mixture/plasma studied in greatest detail (100 W, 17.5 % Ar in H$_2$, 300 mTorr), show a significant contribution of cold and warm hydrogen. In particular, at Position 2 (between the electrodes) there is a significant amount of 'warm' hydrogen. However, over time ('late,' >10 hours) hot hydrogen dominates at all three positions.

Some examples of the Balmer α line, fitted and unfitted, for both Ar/H$_2$ and H$_2$ only plasmas, are shown in Figure 3. Note, that in this figure and subsequent figures and tables, unless specifically noted, the data at Positions 1 and 3 was taken with a 30° tilt applied to the light fiber. It is also important to note that the line broadening is not limited to the H$_\alpha$ line, but is present in about the same magnitude (energy terms) in the entire Balmer series, which allowed the excitation temperature to be measured at various times and locations (more later).

It is also clear that the temperature of the hot H in the Ar/H$_2$ plasmas is only a weak function of position and time (Figure 4) at 100 W. In contrast, the temperature at higher powers in mixed gas plasmas and in pure H$_2$ plasmas at all powers is a fairly strong function of position within the plasma (Figure 5). In many cases for the mixed gas plasma the highest H atom temperature is found at Positions 3. Clearly, the average temperature of the H atoms does not correlate with the accelerating voltage in this case. However, in all cases for pure H$_2$ plasmas the highest temperature is at Position 2, between the electrodes (Figure 5).

At 100 W the operating parameter that has the strongest influence on the temperature of the hot H is the fraction of Ar in the system. As shown in Figure 6 the measured energy of the hot H atoms in the mixed gas plasma at Position 3 is a strong function of the argon fraction, actually reaching a peak value at about 5% argon. The relationship between hottest hydrogen and gas mixture composition is not simple, suggesting 'second order effects.' Indeed, in pure H plasmas the hottest atomic hydrogen is generally found at Position 2 (see Figure 5), and in some cases is hotter than any hot H temperature measured in any mixed gas plasma.

In earlier studies of $H_2$/He (29 ± 3eV) or pure water plasmas (>40 eV) run in the same system at 100 W [12, 14], no evidence was found of a significant effect of position or power on hot H temperature. Clearly, the identity of the gas paired with hydrogen is critical.

It was also found that the concentration profiles of hot H were not consistent with a field production mechanism. As shown in Figure 7, under some conditions the highest hot H concentration was found at the gas entry point, and the lowest concentration at the gas exit. In contrast, the concentration profiles of all the other species studied, cold H, warm H and excited argon, peaked between the electrodes, suggesting that the mechanisms for producing those species is enhanced by the field or by some species found in greatest quantity in the highest-field or power-dissipation region.

**B. Angular Dependence**

Figure 8 shows the angular profile of the Doppler energy of hot hydrogen at position 1 and position 3 wherein the reference is normal to the plasma axis. As shown in Figure 8, there is little impact of the tilt angle, either at Position 1 or Position 3, on the measured broadening. It is also clear that the observed Doppler broadening is symmetric at all angles.

### B. Field Parallel and Field Perpendicular Dependence

There is no measurable impact of the orientation of the observation relative to the field direction on the magnitude of broadening or symmetry of the Balmer series lines. As shown in Figure 9, the Doppler lines at observation Position 2 (perpendicular to the field) and Position 2b (parallel to the field) are both totally symmetric. This was found to be true over a broad range of compositions and powers. In Table V, the magnitude of the broadening at Positions 2 and 2b is shown to be virtually identical at any specified set of operating conditions.

### C. Temporal Behavior

One particular mixture, 17.5 % Ar/82.5% $H_2$ (volume), was studied for its temporal behavior. It is notable that there are changes in the plasma over time. The most significant change was the fraction of H atoms in the 'hot' state gradually increases over time until it is more than 95% at all locations (Figure 10). Another notable change is the ratio of the Ar intensity to that of atomic hydrogen. As shown in Figure 10b the total Balmer $\alpha$ signal relative to the Ar 696 nm line decreases over time. This correlates with an increase in the intensity of the excited Ar signal as the atomic hydrogen signal change little, approximately 30 %, over the duration of the experiments. Other measurable plasma characteristics, including excitation energy and electron temperature are virtually unchanged over time.

### D. Temperature Probes

Both the electron temperature and the excitation temperature were measured at all three positions as a function of time in the 17.5% Ar/82.5% $H_2$ plasma. The excitation temperature, computed from the relative intensity of the Balmer series lines, was found to be below 1 eV at all times and at all positions. It was found to be a weak function of position, always somewhat higher away from the electrodes (Figure 11). Average electron temperature was measured as a function of position (see Table IV for details) and was found to be slightly higher between the electrodes, but less than 0.8 eV everywhere in the cell. It is notable that the magnitude of the two temperatures, although not the temperature profiles, are very similar. This suggests that all the species within the plasma are thermalized, except the hot H atoms.

Control Mixtures- In earlier studies control plasmas, consisting of $H_2$/Xe mixtures were shown to produce no 'hot' hydrogen in conditions in which the cell was not metallized (early). In the region between the electrodes (Position 2) in those earlier studies, 'warm' (ca. 5 eV) hydrogen was found, but at the ends of the cell (Positions 1 and 3) only cold hydrogen was found. In the present work, the Xe concentration was lower than in the earlier study, and we observed changes in the H atom signal as a function of time (Figure 12).

One other feature of the control studies should be noted: the concentration of H atoms in the Ar/$H_2$ plasmas was always far greater than that observed in the Xe/$H_2$ plasmas, moreover, in the Xe plasmas the H atom signal strength drops significantly with time, whereas in the other plasmas it increased slightly with time. That is, comparisons of the atomic hydrogen ($H_\alpha$) signal strength were made at Positions 2 and 3 between the Xe/$H_2$ plasma and the Ar/$H_2$ plasma in which the fraction noble gas (17.5%), and all other operating conditions, varied over a range of power (100, 150 and 200 W) and pressures (80 and 300 mTorr), were the same. In all of these studies the Ar/$H_2$ plasma always generated an H atom signal far greater than that obtained in the equivalent Xe/$H_2$ plasma. Indeed, the ratio of the H signal strength in Ar/$H_2$ to that in Xe/$H_2$ varied between 15 and 150.

## IV. DISCUSSION

This study contains several results that profoundly challenge the earlier proposed models for selective line broadening in $H_2$/Ar plasmas. It is also the third GEC line broadening study that contains data fully consistent with the Mills' model of energy production via the chemical/catalytic transition of the lone electron in a hydrogen atom to a sub-traditional ground state.

In our earlier papers we discussed in detail the various models of line broadening presented in earlier $H_2$/Ar plasma studies in which selective H atom line broadening was observed. In all variations on those models one consistent component is that the energy required for the selective heating of the hydrogen atoms is directly absorbed by ions from the field. Yet, recent data appears to have persuaded some proponents of this model to propose changes to virtually every aspect of the model, including this one. Indeed, the most recent modification of the model has a new name: 'collision model' (CM). Consistent with this appellation, in the modified model there is no longer a need for ions to absorb energy from a field at all. Rather, the modified theory merely requires that in the low field regions of the plasma 'fast electrons' are present. Remarkably, in these regions (e.g. negative glow region) the fast electrons produce hot atomic hydrogen of an energy virtually indistinguishable in magnitude from that recorded in the high field region. The mechanism by which this energy is selectively provided to atomic hydrogen, and not for example to molecular hydrogen, is not described. Moreover, the CM model for line broadening is a function of position within the plasma. That is, the proponents of the CM model do not find any inconsistency in the suggestion that the mechanism of generating hot hydrogen in the low field region is completely different from that in the high field regions and the transition is seamless and independent of plasma parameters. That is, in the high field regions of the plasma the CM model postulates direct acceleration of hydrogen ions is responsible, as per earlier versions of the model, for providing the energy required to produce the observed Doppler broadening.

The CM model also no longer requires hot hydrogen to bounce off surfaces (e.g. electrodes) in order to explain the observation, clearly found in the data of other workers [23], as well as herein, that the observed Doppler broadening is symmetric from all observation orientations relative to the field, as well as in plasma regions in which there is no field. In the CM model it is proposed that a Gaussian (directionally random) distribution is achieved via "collision excitation of $H_f$ on $H_2$ with large angle scattering", where $H_f$ are hot hydrogen atoms that have previously collided with the electrode. The implications of this process on the conservation of angular momentum is not discussed, nor is the mechanism of extension of this process to regions far from the electrode where presumably any $H_f$ species have equilibrated with the plasma via multiple collisions. Nor, is the crucial role of the gas such as helium or argon versus neon, krypton, or xenon discussed [7, 15-16].

Such models cannot explain the existence of hot hydrogen up to 15 cm from the electrode in a 500 mTorr plasma. First, there is virtually no field in this region. Indeed, outside the sheath region of a RF plasma it is generally understood that the field is quite weak [10]. Data regarding electron temperature in the present work support this general understanding. These temperatures were found to be less than 0.7 eV throughout the region away from the electrodes. This belies the suggestion that ions of sufficient energy (>15 eV), that is with energy more than an order of magnitude hotter than that measured for the electrons, could be produced by the field present 15 cm from the electrodes. Indeed, outside of the cathode fall region there is no mechanism for producing ions hotter than the far more mobile, hence more easily heated by the field, electrons.

Also, the measured electron temperatures are very close to the 'excitation temperature' measured for this plasma, suggesting that (except for the H atoms) the plasma outside the electrode region is thermalized. There is no reason to believe that there is a significant population of super hot electrons in the tail of the distribution in this region. Nor is there any plausible explanation/mechanism for the suggestion contained in the CM model that these electrons selectively heat hydrogen atoms.

It is also clear that no 'field heating' mechanism can explain the gradient in hot H atoms found in some cases. That is, at 150 W and above absorbed power, the hot H signal in the mixed gas plasma is higher at the end of the cell than it is between the electrodes. If hot H is generated in the high field region between the electrodes, it must then diffuse against the concentration gradient in order for the signal to be stronger 15 cm 'upstream' from its point of origin.

Other mechanisms for 'field heating' of any species leading to the generation of hot atomic hydrogen seem extremely unlikely. For example, could ions formed in the cathode fall region 'escape,' travel 15 cm, and then create hot hydrogen atoms? This postulate requires two unphysical processes. First, the excited ions are required to escape the cathode, second, they are required to travel 15 cm through a gas at ca. 500 mTorr without any energy loss. Even a single elastic collision would partially thermalize a hot ion. Can any model that assumes the initial source of energy for generating hot neutral atomic hydrogen is energy absorption from the field by ions explain the following observations: (i) nearly constant temperature hot atomic hydrogen, more than 20 times greater than that of the electrons, is found throughout the plasma, (ii) the excitation energy of the hydrogen, throughout the plasma, is about 20 times less than its kinetic energy.

We attempted to answer the above question by considering a variety of possible scenarios. For example, we considered the possibility that hot hydrogen atoms generated by collisions of hot ions with surface hydrogen species, traveled 15 cm without loosing energy. This scenario was rejected because the mean free path of <1 cm is not consistent with such a suggestion. We considered the possibility that the field away from the electrodes was far higher than the mV/cm anticipated. Field screening by the sheath reduces the fields dramatically within millimeters, and a highly conductive plasma bulk is essentially equipotential [10]. We rejected this because that would require a new understanding of plasmas, Maxwell's equations, etc. And it would result in far hotter electrons than observed. The sum of our efforts was unsuccessful. We were unable to imagine a single reasonable scenario in which 'hot ions acceleration in the field' can explain the existence of >15 eV H atoms throughout the plasma.

In RF discharges, the atomic hydrogen intensity is greatest at the electrode surface which is indicative of the high atomic hydrogen concentration present here. As the only explanation found by Mills et al. for the broadening of the hydrogen lines in pure hydrogen microwave plasmas, the role of two hydrogen atoms as a catalyst for a third hydrogen atom to form the hydrino H(1/2) was discussed previously [15-16]. Ordinary hydrogen atom undergoing a catalysis step to $n=\frac{1}{2}$ releases a net of $40.8 \text{ eV}$. Since each has a ionization energy of $m \cdot 27.2 \text{ eV}$, hydrinos may serve as catalysts as discussed previously [28-30]. Thus, further catalytic transitions may occur: $n=\frac{1}{2} \to \frac{1}{3}, \frac{1}{3} \to \frac{1}{4}, \frac{1}{4} \to \frac{1}{5}$, and so on by the autocatalysis of hydrinos. The high concentrations of hydrogen atoms at the cathode of RF and glow discharges are favorable for these reactions to proceed.

The observation of an elevated hydrogen atom temperature for pure hydrogen plasmas and mixtures containing hydrogen with the unusual absence of an elevated temperature of any other gas present has also been observed before for RF and glow discharges. For example, using a GEC RF cell Radovanov et al. [4] observed that the structure of the $H_\alpha$ line emission from a pure $H_2$ discharge showed a slow component with an average energy of 0.2 eV and a broadened component of 8.0 eV. Very high energies have been observed. Hydrogen line broadening corresponding to 123 eV has been observed with hydrogen plasmas maintained in a GEC RF cell [10]. Extraordinary line broadening near the cathode corresponding to fast H with >300 eV is only observed in the cases of discharges of hydrogen or in hydrogen mixtures. This phenomenon is not observed in discharges of pure noble gases [3, 5, 10, 31-33]. In the case of the production of fast H, the intensity may be low due to efficient collisional energy exchange with molecular hydrogen with dissociative excitation [34]. In a glow discharge, fast H is formed and excited predominantly near the electrode surfaces. Under high pressure conditions and short duration runs, the emission from fast H formed at the cathode is also not expected to extend significantly into the bulk of an $H_2$ discharge because of quenching of $H(n = 3)$ by collisions with $H_2$ [4].

Again, this unusual effect was attributed to electric field acceleration of positive hydrogen ions in the cathode fall region. In our RF hydrogen plasmas, we have further shown that the broadening exits where no such field exists, but the potential for an rt-plasmas does. Since the ionization energy of hydrogen is 13.6 eV, two hydrogen atoms can provide a net enthalpy equal to the potential energy of the hydrogen atom, 27.2 eV—the necessary resonance energy, for a third hydrogen atom to form H(1/2).

$$2H + H \rightarrow H(1/2) + 2H + 40.8 \text{ eV} \tag{1}$$

On this basis with subsequent highly energetic transitions to further lower-energy states with part of the energy channeled to fast H production, the unusual observation of the H energy up to one hundred times above the electron temperature is expected. The rt-plasma mechanism is also consistent with the observation that the broadening increases with time in the non-field regions. The effect is expected to be more pronounced with greater hydrogen concentration such as that achieved near or on the cathode in RF and glow discharge cells. Past studies have shown the importance of the surface conditions that permit a high atomic hydrogen concentration at the cathode [5]. But, once formed subsequent transitions of H(1/2) to lower states by hydrinos yields very high energy release in extra-cathode regions where there is essentially no electric field [29]. The production of fast H due to this energy release that is independent of the field or proximity to the cathode surface is consistent with our observations.

High densities of atomic hydrogen corresponding to one or more monolayers are easily formed on metal surfaces [35-38]. Two atoms of atomic hydrogen may serve as a catalyst for a third atom (Eq. (1)) with the highest rate on or near the surface of the electrode. The product, gaseous hydrino atoms, may then react to lower-energy states in the gas phase throughout the cell. Since the products accumulate over time and the reactions to lower states are exponentially energetic, broadening is predicted away from the electrodes. Furthermore, the broadening is anticipated to increase in intensity and energy with time until a steady state is reached. These predictions were observed in our experiments with hydrogen alone and in previous experiments with water vapor [39].

It is known that water vapor increases the hydrogen concentration of hydrogen plasmas as shown by Kikuchi et al. [40]. The atomic hydrogen concentration is also increased with the addition of oxygen as shown by Chabert at al. [41]. In addition to the potential for oxygen to act as a catalyst [14, 39], a water vapor plasma is also anticipated to show extraordinary broadening that is time dependent, but position independent due to H catalysis by 2H. These predictions were also observed as reported [14, 39].

Below we show that there are several mechanisms from Mills CQM that would be consistent with the results obtained with pure $H_2$ and $Ar/H_2$ plasmas. In the CQM model, it is postulated that the electron in the hydrogen atoms that undergo a 'catalytic' reaction with $Ar^+$, or with an existing hydrino, move from the 'traditional' ground state (principle quantum number, $n=1$) to a 'fractional' quantum state (e.g. $n=1/2$). Thus, the requirements of any mechanism consistent with CQM are: (i) energy released be of a value consistent with that expected by the model of Ar ion catalysis, (ii) that the process would conserve momentum, and (iii) fast atomic H (not just a hydrino) is created. Some explanation for the other 'non-equilibrium feature' of the plasmas, the existence of an extremely high H atom population, is also needed.

In the CQM model several two body processes are consistent with these requirements: two existing hydrinos, collide to create a hot hydrino and a hot H atom. For example, an H(1/2) and and an H(1/2) hydrino upon collision can produce an H atom and an H(1/3) hydrino:

$$H(1/2) + H(1/2) \rightarrow H + H(1/3) + 27.2 \text{ eV} \qquad (2)$$

Given the requirement that momentum be conserved, each product 'atom' would gain half, or 13.6 eV from this process. That is, the two equal mass products would need to go in 'opposite' directions, with equal kinetic energy, in order for no net momentum to be generated by the process. Naturally, only the (subsequently) excited H atoms produced by this process would produce visible light.

An example of a relevant two-body collision, predicted to occur by the CQM model, that could only indirectly produce hot H would be the collision between H and $Ar^+$:

$$H + Ar^+ \rightarrow H(1/2) + Ar^{++} + 13.6 \text{ eV} \tag{3}$$

It is 'resonant transfer processes' [15,16], of this type, key postulates in the CQM model that initialize the generation of hot H atoms. In this particular reaction, the conservation of momentum requirement would yield a 'hot' hydrino with about 13 eV of kinetic energy. However, hot hydrinos do not produce Balmer series light. On the other hand, a hot hydrino, colliding with a hydrogen molecule would easily provide enough energy to break the molecular bond and still generate hot H atoms. In this study, in which the majority of species are always H or $H_2$, this is quite plausible. Moreover, co-consideration of energy and momentum conservation indicates that a hot hydrino colliding with an argon atom would only loose 5% of its energy. Clearly, it would take many collisions with argon species to 'thermalize' a hot hydrino, allowing for many collisions, in a majority hydrogen plasma, with molecular hydrogen that could produce hot H atoms.

Processes such as the proposed catalytic reaction that occur without photons and that require collisions or nonradiative energy transfer are common. For example, the exothermic chemical reaction of H + H to form $H_2$ does not occur with the emission of a photon [42]. Rather, the reaction requires a collision with a third body, $M$, to remove the bond energy

$$H + H + M \rightarrow H_2 + M^* \tag{4}$$

The third body distributes the energy from the exothermic reaction, and the end result is the $H_2$ molecule and an increase in the temperature of the system. Some commercial phosphors are based on nonradiative energy transfer involving multipole coupling. For example, the strong absorption strength of $Sb^{3+}$ ions along with the efficient nonradiative transfer of excitation from $Sb^{3+}$ to $Mn^{2+}$, are responsible for the strong manganese luminescence from phosphors containing these ions [43].

Reactions (2) and (3) are only two specific examples of a postulated general 'resonant transfer' process that leads to the release of energy when H atoms collide with catalytic species, according to the Mills' model. That is, the reactions above are examples of a catalytic resonant energy transfer from hydrogen atoms to $Ar^+$ or $He^+$ ions during a catalytic collisional event. The magnitude of the broadening observed is consistent with the magnitude of the energies of the initial and subsequent catalytic reactions that may be transferred to form fast H [28-30, 44].

As noted in earlier Mills' papers, the 'resonant energy' catalysis reaction can be written in more specific terms. For example, for $He^+$,

$$54.417\ eV + He^+ + H[a_H] \rightarrow He^{2+} + e^- + H\left[\frac{a_H}{3}\right] + 108.8\ eV \tag{5}$$

$$He^{2+} + e^- \rightarrow He^+ + 54.417\ eV \tag{6}$$

And, the overall reaction is

$$H[a_H] \rightarrow H\left[\frac{a_H}{3}\right] + 54.4\ eV + 54.4\ eV \tag{7}$$

Since the products of the catalysis reaction have binding energies of $m \cdot 27.2$ eV, they may further serve as catalysts. Thus, further catalytic transitions may occur:

$$n = \frac{1}{3} \rightarrow \frac{1}{4},\ \frac{1}{4} \rightarrow \frac{1}{5} \dots \frac{1}{137}.$$

Since the reactions given by Eqs. (5-7) involve two steps of energy release, it may be written as follows:

$$54.417\ eV + He^+ + H[a_H] \rightarrow He^{2+} + e^- + H*\left[\frac{a_H}{3}\right] + 54.4\ eV \tag{8a}$$

$$H*\left[\frac{a_H}{3}\right] \rightarrow H\left[\frac{a_H}{3}\right] + 54.4\ eV \tag{8b}$$

$$He^{2+} + e^- \rightarrow He^+ + 54.417\ eV \tag{8c}$$

And, the overall reaction is

$$H[a_H] \rightarrow H\left[\frac{a_H}{3}\right] + 54.4 \ eV + 54.4 \ eV \tag{9}$$

wherein $H^*\left[\frac{a_H}{3}\right]$ has the radius of the hydrogen atom and a central field equivalent to 3 times that of a proton and $H\left[\frac{a_H}{3}\right]$ is the corresponding stable state with the radius of 1/3 that of H.

H is the lightest atom; thus, it is the most probable fast species in collisional energy exchange from the H intermediate (e.g. $H^*\left[\frac{a_H}{3}\right]$). Additionally, H is unique with regard to $H^*\left[\frac{a_H}{p}\right]$ in that all are these species are energy states of hydrogen with corresponding harmonic de Broglie wavelengths. Thus, the cross section for H excitation by a through-space nonradiative energy transfer to form fast H is predicted to be large since it is a resonant process. Efficient energy transfer can occur by through-space mechanisms that are common such as dipole-dipole interactions as described by Förster's theory [45–49]. The hydrino transition energy transferred to form hot hydrogen that undergoes collisional excitation can also be resonant since the energy transfer is $q \times 13.6$ eV. The Frank Hertz experiment is an example of resonant-collisional excitation. This may also be the mechanism of achieving highly excited (pumped) states in water-vapor plasmas reported previously [13, 50].

Consequently, rather than radiation, a resonant transfer to form fast H may occur, and the emission of H($n = 3$) from fast H($n = 1$) atoms excited by collisions with the background $H_2$ may occur as discussed previously by Radovanov et al. [10]. Since additional catalytic reactions of varying energies are possible as discussed previously [28-30], the particular conditions in the cell may favor more than one H population. Collisional energy transfer between fast H and matrix gas may also give rise to a bimodal or trimodal distribution.

The thermalization of the energy transferred to the catalyst may contribute to the formation of fast H. As discussed *supra.*, excessive line broadening was observed in the case where $Ar^+$ was present with hydrogen, but not when xenon replaced argon as predicted since argon ion may form an rt-plasma [15-19, 51]. It provides a net positive enthalpy of reaction of 27.2 eV (i.e. it resonantly accepts the nonradiative energy transfer from hydrogen atoms and releases the energy to the surroundings which heat up). The total energy of the catalysis reaction given previously [28] is 40.8 eV. Thermalization of the 27.2 eV transferred from the catalyst and the 13.6 eV from H (See reaction Eqs. 9-11 of Ref [44]) is consistent with the observation by Djurovic and Roberts [1] and Radovanov et al. [10] of no directional effects of the Doppler broadening due to the applied electric field and the average energy of 23.8 eV and 28 eV, respectively, of the fast H excited atoms (> 10 eV excitation energy) that was similar throughout the whole interelectrode region of the discharge over a wide range of gas pressures, applied RF voltages, and hydrogen concentration in $Ar/H_2$ mixtures. In addition, at low pressures, Radovanov et al. [4] observed $Ar^+$ and $ArH^+$ kinetic energy distribution profiles with an edge at about 27.2 eV. Additional evidence for the formation of $Ar^+$ with kinetic energies of $q \times 13.6$ eV was reported previously [52]. Further studies with xenon would provide a test of the proposed catalytic rt-reaction.

The general model described above explains many specific observations made in this study. It can explain the fact that the amount of H observed in $H_2/Ar$ plasmas in the region away from the electrodes in some cases is higher than that observed between the electrodes. It can also explain why virtually all the observed H is fast H in some cases with the energy from the reaction matching that of the hot H. Regarding the former, the temperature of the electrons observed in the region away from the electrodes is below that required to split $H_2$, hence a low concentration of H is expected. Only electrons in the tail of the EEDF will have any ability to create atomic hydrogen. Given the fact that there is no evidence of hotter electrons in the mixed gas plasma, the relative high concentration observed away from the electrodes in pure hydrogen and $Ar/H_2$ mixed gas plasmas requires a 'non-electron' mechanism for atomic H generation.

The mechanisms we outlined above, such as hydrino-hydrino collisions, but also hydrino-$H_2$ molecule collisions, provide the additional needed routes to H atom production:

$$H_{very\ hot} + H_2 \rightarrow 3\ H_{hot} \tag{10}$$

In some fashion the above postulated reaction is similar to those proposed in the CM model described earlier. However, unlike the CM model according to CQM this process is not field dependent, nor does it require 'hot electrons', thus it can and should be effective everywhere within the plasma. Naturally, as this process will take place with both a conservation of energy and momentum, hot hydrogen will have isotropic directionality and a Gaussian distribution of energies.

Note, once more H atoms are generated, the 'kinetics' for additional hydrino creation are improved. That is, if we assume that only two body collisions are relevant, than hydrino formation can reasonably be modeled as first order in H atom concentration. And, mechanisms of the type suggested can also explain the very high fraction of hot H atoms found. That is, hydrogen atoms produced either via the hydrino-hydrino collision mechanism or the hydrino-molecule mechanism will be 'hot' H.

As outlined above in the general model section, and elsewhere [29-30] the process is autocatalytic. Once hydrinos are formed, they accelerate the production of H atoms, which in turn enhances the rate of hydrino formation via catalytic processes. Moreover, once hydrinos are formed, they can act as catalysts, for the formation of additional hydrinos from H atoms. Thus for example, an H$_{1/2}$ hydrino can interact with an H atom to make two H$_{1/2}$ hydrinos. To wit:

$$H(1/2) + H \rightarrow 2\,H(1/2) + 27.2 \text{ eV} \tag{11}$$

The hot hydrinos produced in this fashion can clearly create additional H atoms by collisions with H$_2$ molecules. An autocatalytic kinetic scheme of this type in which hot hydrinos form, in turn increase the H atom density, and then in turn, are catalytically converted to hydrinos by collisions can clearly produce a steady-state concentration of hydrinos once a few hydrinos are formed.

It is notable that each type of catalyst creates a different predominant hydrino. For example, in the case of He, it is expected that H(1/3) will form, with an excess kinetic energy of 54.4 eV. In contrast, in an Ar/H$_2$ plasma H(1/2) with an excess energy of either 27.2 eV or 13.6 eV will form due to the catalytic action of argon. These differences in the energy from the process, as well as the relative reaction rates and quenching by other gas atoms or molecules, may account for the observation that changing the catalytic gas changes the observed temperature of the hot hydrogen. For example, in our GEC system we have seen distinct differences between predominant helium He/H$_2$ plasmas (about 30 eV), H$_2$O (greater than 40 eV), predominantly hydrogen Ar/H$_2$ plasmas (ca. 20 eV) and pure H$_2$ plasmas (initially 13 eV and later ca. 55 eV). In sum, it is clear that the Mills' model is consistent with observations in this and earlier studies by our group that the precise nature of the hot H population is a function of global gas composition, and even local gas composition. Regarding the latter, it is reasonable to postulate that the populations of H atoms, molecules and hot hydrinos can vary significantly between points between the electrodes, at the entry, and at the exit to the GEC-type cell to an extent sufficient to dramatically impact the nature of the predominant reaction pathways at those positions. This in turn will result in a modification of the hot H populations locally. It is also clear that the gas composition impacts hot hydrogen energy far more than changes in the accelerating voltage.

One other feature of the present data consistent with the CQM model is the failure to observe any 'hot' hydrogen in $H_2$/Xe plasmas. According to this model 2H, $He^+$, $Ar^+$ and 2O can act as 'catalysts' for the process. It is notable in this regard that in earlier studies we have shown selective line broadening of atomic hydrogen in $H_2$/He plasmas ($He^+$ catalyst) and water plasmas (O catalyst). In contrast, no Xe species is capable of acting as a catalyst. The low concentration of atomic hydrogen in any form (Figure 12) in the Xe/$H_2$ plasma is also consistent with the model. That is, in the absence of hot H atoms produced via the mechanisms postulated above, there are virtually no species, other than a few electrons in the tail of the EEDF, present in the plasma with sufficient energy to dissociate $H_2$. Hence, one prediction of the CQM model of line broadening is that in the absence of sufficient catalytic gas, not only will there be no hot H atoms, the concentration of H atoms of all types will be dramatically reduced.

As noted above, many spectroscopic studies are consistent with the Mills' hypothesis, including the observation of lines at energies specified by the theory in the EUV part of the spectrum [28-30, 53]. It is also clear that non-spectroscopic data is also consistent with the Mills' hypothesis. An exhaustive set of calorimetric studies was conducted and the results were completely consistent with the model, and in contrast have no explanation in conventional QED physics. For example, eleven types of control plasmas (e.g. Kr, Kr/$H_2$, $N_2$, $N_2$/$H_2$), many run repeatedly, all produced 37.5 +/- 2.5 W of energy, as expected from an independent measure (Agilent power diodes) of the energy fed to these plasmas, according to a water bath measurement. In contrast, three types of predicted resonant transfer plasmas, pure water, Ar/$H_2$, and He/$H_2$, in the identical system, using the same operating conditions of flow, pressure, and the same measured power from the microwave sources as the controls, repeatedly produced far more. For the He/$H_2$ plasmas, around 50 W was observed, between 55 and 60 W was observed for the Ar/$H_2$ plasmas, and repeatedly more than 65 W of power was found in output to the water bath from the pure water plasmas [54].

Regardless of the merits of the Mills' model, the present data challenges our fundamental understanding of selective line broadening of H atoms in gas plasmas. That is, at present, it is clear there is no 'traditional physics' model capable of explaining the selective heating of H atoms in low field regions of $H_2$ only and $H_2$/Ar plasmas.

A final value of the present work is the unexpected bounty of information regarding the impact of a wide variety of operating parameters on the characteristics of the hot H populations. It is clear that this work, in which more than 1500 spectra were collected and analyzed, still does not provide a complete picture of the phenomenon. A wider range of gas compositions, pressures, and powers should be studied. More thorough mapping might prove valuable. Mapping the influence of parameters not studied at all, such as flow rates, could provide valuable insights. We would like to collect information about the behavior of the plasma immediately after it is turned on and off. The use of deuterium might provide interesting information. Additional characterization methods, such as calorimetry would certainly be of value.

It should be noted that the GEC-type cell approach to detection of selective Balmer $H_\alpha$ line broadening is very robust. Thus, those intending to replicate this work will find the task relatively straightforward, with one caveat. That is, it is quite clear that operating conditions do impact the results. Hence replication requires adhering relatively closely to the operating conditions described herein.

Table I. Doppler energy (temperature) of hydrogen and percentage of peak areas from Balmer $H_\alpha$ line collected in Position 1. (Plasma gas 82.5 % $H_2$/17.5 % Ar, Pressure 0.28 Torr, 100 W RF Power.)

| Time | Temperature[a] | | | Percentage of Peak Area | | |
|---|---|---|---|---|---|---|
| | Cold | Warm | Hot | Cold | Warm | Hot |
| (hours) | (eV) | (eV) | (eV) | | | |
| 0 | 0.12 | 1.35 | 20.13 | 9% | 24% | 67% |
| 3 | 0.15 | 3.42 | 22.59 | 14% | 27% | 59% |
| 4 | 0.12 | 0.91 | 14.08 | 7% | 4% | 89% |
| 5.5 | 0.12 | 1.59 | 14.45 | 7% | 5% | 88% |
| 7 | 0.12 | 0.91 | 14.99 | 7% | 3% | 90% |
| 11 | 0.11 | 0.38 | 17.85 | 3% | 2% | 95% |
| 16 | 0.12 | 1.05 | 16.93 | 5% | 10% | 85% |
| 31 | 0.09 | 0.53 | 15.97 | 2% | 7% | 91% |
| 34 | 0.07 | 0.38 | 19.84 | 1% | 4% | 95% |
| 41 | 0.22 | xx[b] | 16.21 | 1% | 0% | 99% |

[a] The temperatures were calculated using the following equation: $\Delta\lambda_{FWHM} \approx 7.1626 \times 10^{-7} \lambda(\text{nm})\sqrt{T(^\circ K)}$
[b] No warm hydrogen detected

Table II. Doppler energy (temperature) of hydrogen and percentage of peak areas from Balmer $H_\alpha$ line collected in Position 2. (Plasma gas 82.5 % $H_2$/17.5 % Ar, Pressure 0.28 Torr, 100 W RF Power.)

| Time | Temperature | | | Percentage of Peak Area | | |
|---|---|---|---|---|---|---|
| | Cold | Warm | Hot | Cold | Warm | Hot |
| (hours) | (eV) | (eV) | (eV) | | | |
| 0 | 0.15 | 1.40 | 18.46 | 4% | 51% | 45% |
| 3 | 0.11 | 0.71 | 11.00 | 5% | 6% | 89% |
| 4 | 0.05 | 0.10 | 11.74 | 1% | 2% | 97% |
| 5.5 | 0.10 | 1.56 | 12.14 | 3% | 4% | 93% |
| 7 | 0.13 | 1.38 | 14.68 | 4% | 3% | 93% |
| 11 | 0.13 | 0.56 | 14.76 | 8% | 3% | 89% |
| 16 | 0.22 | 3.08 | 16.06 | 6% | 10% | 84% |
| 31 | 0.15 | 1.35 | 13.48 | 4% | 26% | 70% |
| 34 | 0.20 | xx[a] | 14.14 | 2% | 0% | 98% |
| 41 | 0.21 | xx | 13.46 | 2% | 0% | 98% |

[a] No warm hydrogen detected

Table III. Doppler energy (temperature) of hydrogen and percentage of peak areas from Balmer $H_\alpha$ line collected in Position 3. (Plasma gas 82.5 % $H_2$/17.5 % Ar, Pressure 0.28 Torr, 100 W RF Power.)

| Time | Temperature | | | Percentage of Peak Area | | |
| --- | --- | --- | --- | --- | --- | --- |
| | Cold | Warm | Hot | Cold | Warm | Hot |
| (hours) | (eV) | (eV) | (eV) | | | |
| 0 | 0.12 | 1.33 | 18.01 | 11% | 25% | 64% |
| 3 | 0.12 | 1.78 | 17.03 | 14% | 22% | 64% |
| 4 | 0.12 | 1.14 | 14.52 | 10% | 7% | 83 % |
| 5.5 | 0.11 | 1.04 | 16.78 | 18% | 9% | 73% |
| 7 | 0.11 | 0.99 | 15.73 | 15% | 6% | 79% |
| 11 | 0.11 | 0.44 | 15.28 | 5% | 2% | 93% |
| 16 | 0.09 | 0.66 | 14.83 | 5% | 8% | 87% |
| 31 | 0.06 | 0.57 | 14.96 | 2% | 8% | 90% |
| 34 | 0.06 | 0.35 | 14.78 | 2% | 5% | 93% |
| 41 | 0.13 | xx[a] | 14.79 | 2% | 0.00% | 98% |

[a] No warm hydrogen detected

Table IV. Parameters of Ar (430 nm) used for the calculation of electron temperature.

| | |
|---|---|
| $\lambda_L$ | 430 nm |
| $U_i$ | 5.5 |
| $A_{21}$ | $3.1 \times 10^5$ (1/sec) |
| $g_2$ | 5 |
| $G$ | 1.1 |
| $E_i$ | $2.525 \times 10^{-18}$ J |
| $E_2$ | $2.324 \times 10^{-18}$ J |
| $\Delta\lambda$ | 0.013 nm (with the slit 20 μm) |

$$\frac{I_{max}(\lambda)}{I_c(\lambda)} = \left[\frac{h^4 3^{3/2} c^3}{256 \pi^3 e^6 k}\right] \frac{A_{21} g_2}{U_i} \frac{\lambda_L}{\Delta\lambda} \frac{1}{\xi T_e} \exp[\frac{E_i - E_2}{kT_e}]$$

Where:
$A_{21}$ : Einstein transition probability of spontaneous emission between level 2 and 1;
$c$ : Speed of light;
$e$ : charge of electron;
$E_2$ : energy of atom level 2;
$E_i$ : ionization potential;
$G$ : free-free Gaunt factor;
$g_2$ : degeneracy of level 2;
$h$ : Planck's constant;
$I_c$ : observed emission intensity of continuum;
$I_{max}$ : observed emission intensity of the line;
$k$ : Boltzmann constant;
$m$ : mass of electron;
$n_e$ : electron density;
$n_i$ : ion density;
$T_e$ : electron temperature;
$Ui$ : partition function of ion;
$\lambda_c$ : wavelength of continuum;
$\lambda_L$ : wavelength of emission line;
$\Delta\lambda$ wavelength bandwidth;
$\xi$ free-bound continuum correction factor

Table V. Comparison of measured Doppler broadening of hot hydrogen atom and its relative percent population corresponding to optical emission observed perpendicular (Position-2) and parallel and perpendicular (Position-2b) to the electric field for pure hydrogen and 50% Ar/ 50% $H_2$ mixture plasma at different radio frequency power levels.

|  | 200 mT Ar/$H_2$ | | | | 300 mT Ar/$H_2$ | | | | 300 mT $H_2$ | | | |
| --- | --- | --- | --- | --- | --- | --- | --- | --- | --- | --- | --- | --- |
|  | Position 2 | | Position 2(b) | | Position 2 | | Position 2(b) | | Position 2 | | Position 2(b) | |
| Power | $T_\perp$ (eV) | Area (%) | $T_\parallel$ (eV) | Area (%) | $T_\perp$ (eV) | Area (%) | $T_\parallel$ (eV) | Area (%) | $T_\perp$ (eV) | Area (%) | $T_\parallel$ (eV) | Area (%) |
| 100 W | 23.0 | 76.7 | 21.9 | 63.1 | 22.6 | 81.2 | 22.1 | 62.1 | 13.1 | 8.2 | 11.2 | 11.2 |
| 150 W | 23.8 | 78.1 | 23.2 | 63.2 | 24.1 | 79.1 | 22.8 | 64.3 | 13.5 | 10.8 | 11.5 | 13.3 |
| 200 W | 24.1 | 74.2 | 23.9 | 62.7 | 24.9 | 78.6 | 23.9 | 68.4 | 13.5 | 13.5 | 11.9 | 14.4 |

FIG. 1. Schematic of the quartz GEC system. Note that the cell is 14 cm in diameter and 36 cm in length. The cell itself was placed in a grounded blackened aluminum lined box and the fiber optic probe was surrounded by a sheath of grounded aluminum foil as well. Grounded shielding reduced signal noise significantly. A quarter inch hole in the powered electrode allows end-on observation of position-2 parallel to the electric field.

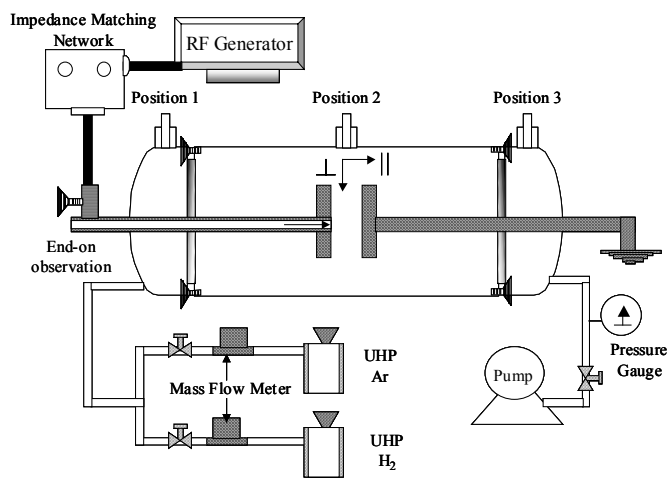

FIG. 2. Peak voltage ($V_p$) measured on the powered electrode at different power levels applied to H$_2$ and 80 % Ar/20% H$_2$ plasma at 500 mTorr.

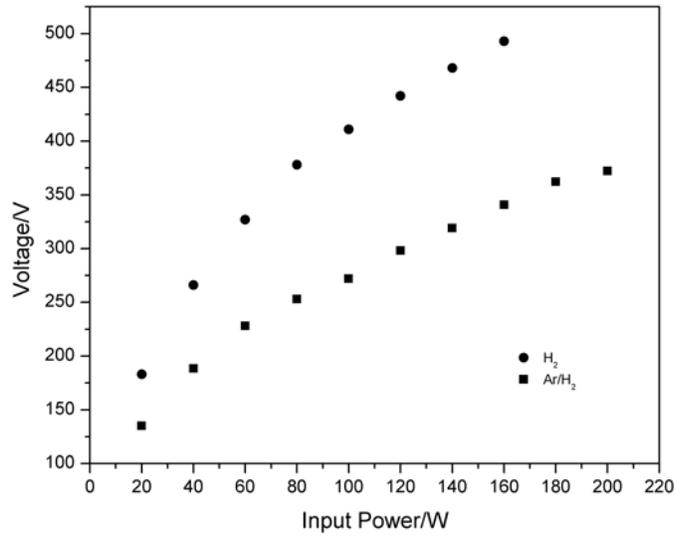

FIG. 3. Raw and fitted H$_\alpha$ lines. (a) Early (<10 hours) stage H$_\alpha$ lines, all positions, of 17.5% Ar/82.5% H$_2$, 100W, 300 mTorr plasma. (b) Fitting of Position 1 data of (a) indicates approximately 65% of signal from hot H. (c) Same as (a) but late stage (>40 hours) H$_\alpha$ lines. (d) Fitting of Position 1 data of (c) indicates approximately 95% of signal from hot H. (e) Low pressure pure hydrogen plasmas produced the hottest H atoms. In this case at 80 mTorr and 100 W, even at Position 3, the hot H temperature was just under 60 eV. The hottest H atoms in mixed gas plasmas under any conditions were never more than 45 eV. In the example shown (100 W, 17.5% Ar/82.5% H$_2$) case the temperature was only 32 eV.

(a)

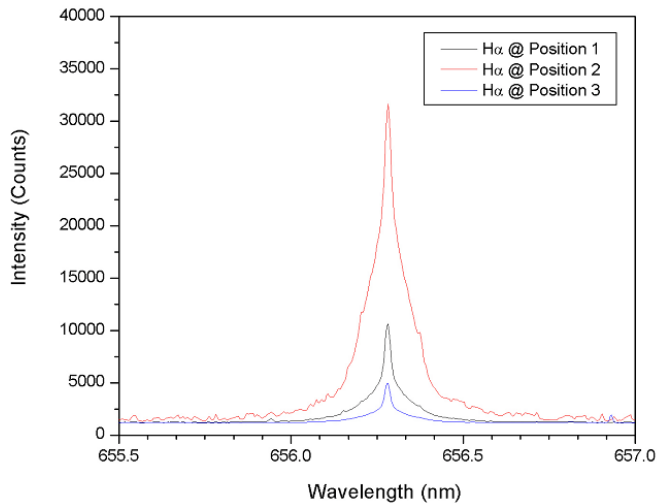

(b)

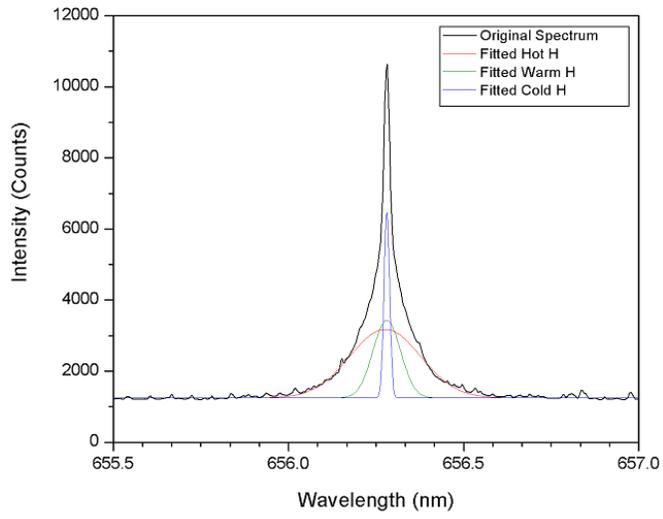

(c)

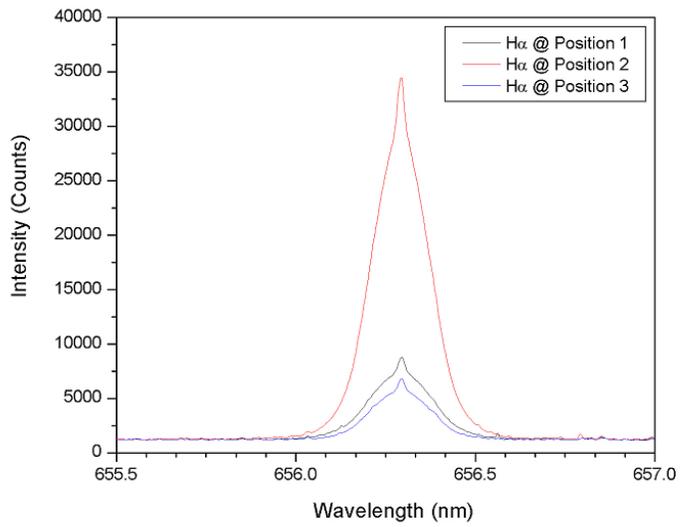

(d)

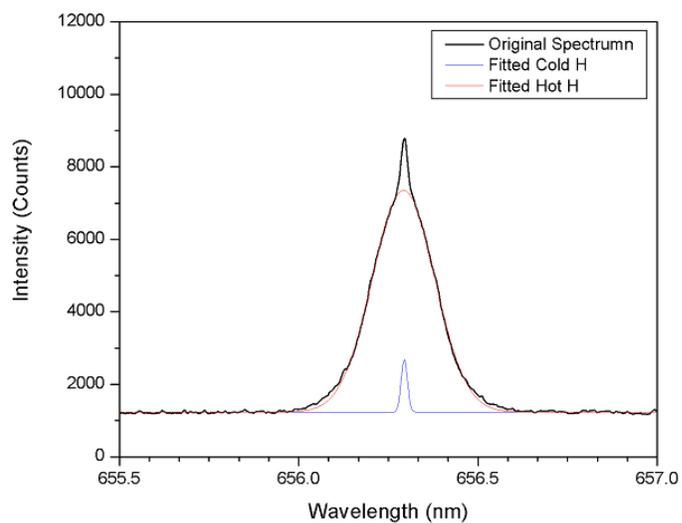

(e)

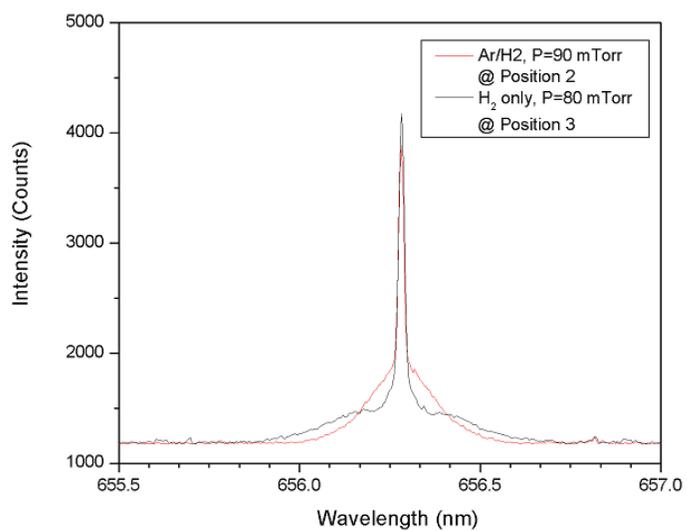

FIG. 4. Impact of time and position on hot atomic H temperature in mixed gas plasma. The measured temperature of hot hydrogen as a function of time for a 100 W 300 mtorr plasma as a function of time and position is shown for a 17.5% Ar/82.5% $H_2$ plasma. The measured value of the hot atomic H temperature is only a weak function of these parameters, but it is notable that the highest temperature H atoms consistently found at the entry (Position 1).

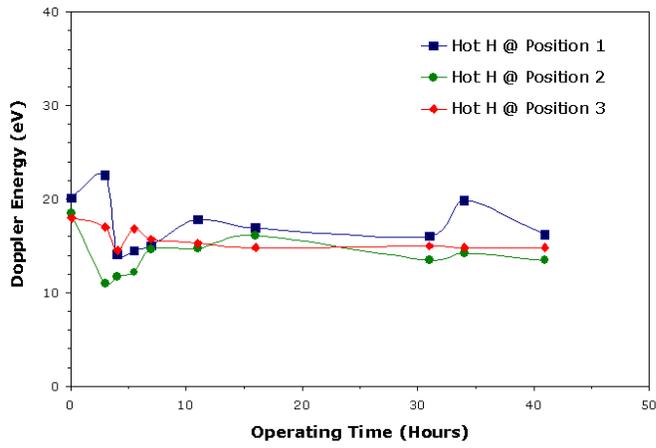

FIG. 5. Hot H temperature as a function of position in Ar/$H_2$ and pure H plasma. (a) Data were taken from 15% Ar/85% $H_2$ plasmas (200 mTorr) 'early' and show that the hot H temperature is a strong function of time and position. The hottest hydrogen in this case is only about 45 eV; (b) In pure hydrogen plasmas (300 mTorr, shown) the hottest hydrogen was always found between the electrodes. At lower pressure (e.g. Figure 3e) hot H temperatures (at all positions) greater than 50 eV are found.

(a)

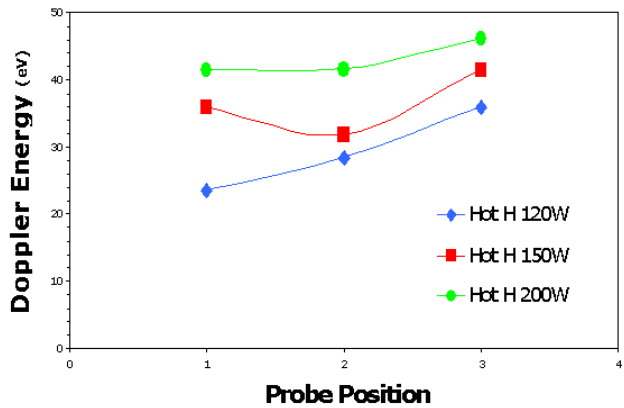

(b)

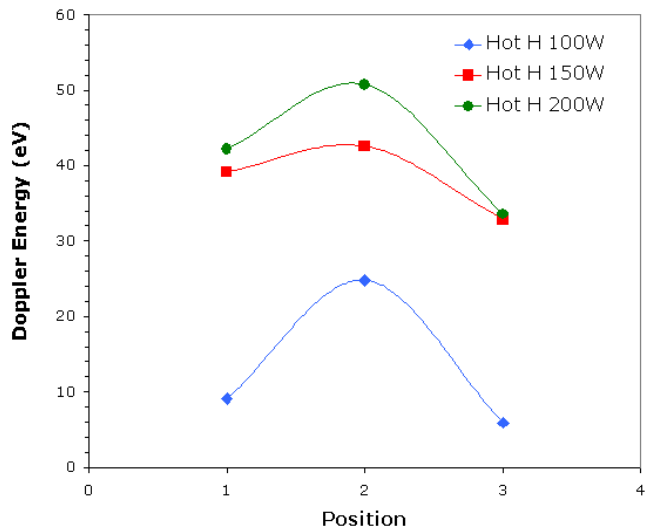

FIG. 6. Hot atomic H temperature as a function of fraction Ar, (semilog scale). (a) Data taken at Position 3 for a series of 300 mTorr, 100 W plasmas show that even a small amount of Ar dramatically reduces the temperature of the hot atomic H. Note that at low pressure, in pure hydrogen plasma, hot H with energy greater than 50 eV can be found at all positions. Also note, relationship is very position dependent (See figure 5). (b) The fraction of hot hydrogen increases with increasing Ar content. The data were collected at Position 3, but the same trend, increasing Ar increased the fraction hot H, was found at all three positions. All data for this figure were taken after the plasmas had been running for more than 20 hours, at constant conditions.

(a)
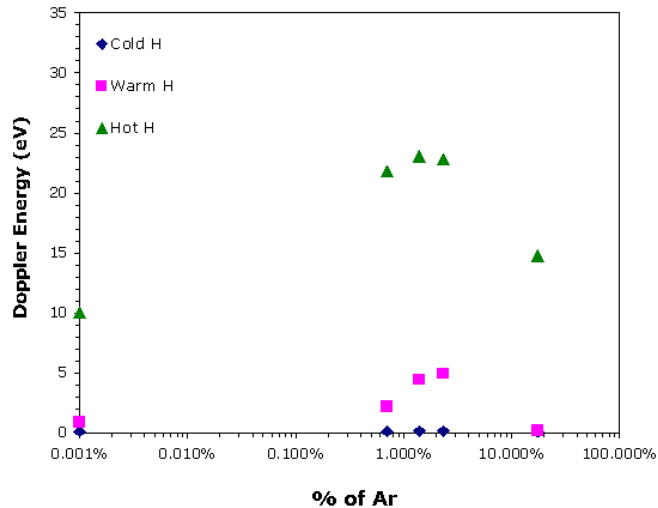

(b)
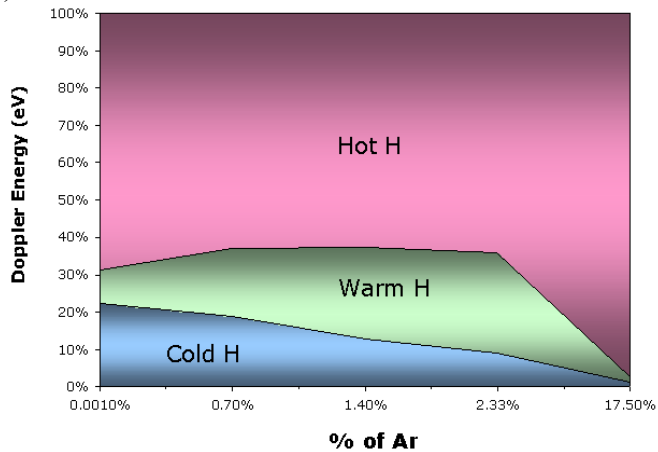

Fig. 7. Relative hot H signal strength. In all cases shown the plasma pressure was 300 mTorr and the composition 17.5 % Ar/82.5% $H_2$. Note that the signal strength of hot H in the 150 and 200 W plasmas are highest at the point at which gas enters the cell (Position 1). In contrast all other species have signal strength profiles that suggest their highest concentrations is between the electrodes.

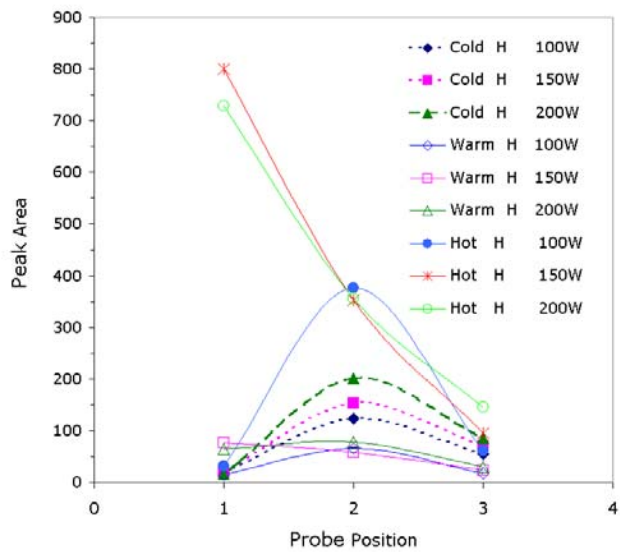

FIG. 8. Angular variation of Doppler energy of hot hydrogen atom in 300 mT 50%Ar/ 50%$H_2$ plasma at 150 and 200 W and 300 mT $H_2$ plasma at 200 W. Plasma emission is collected at Position 1 (Fig. a) and Position 3 (Fig.b). Reference is normal to the chamber axis.

(a)

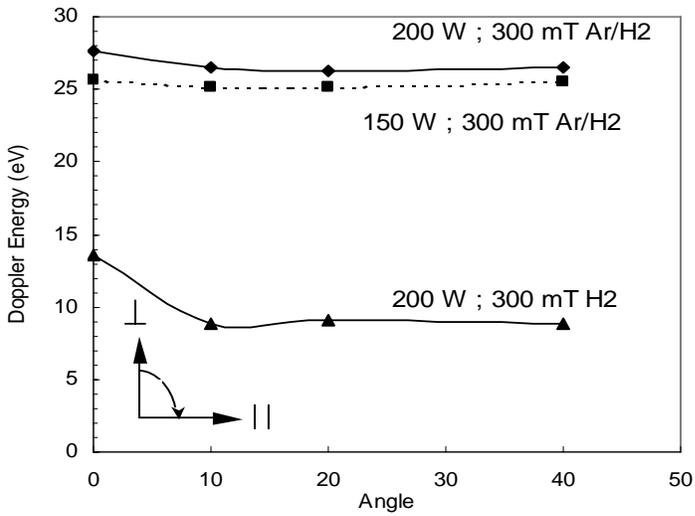

(b)

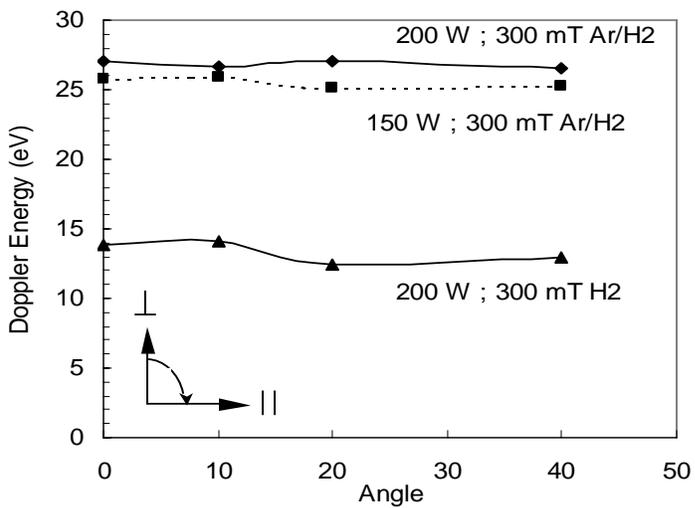

FIG. 9. The graph shows the Doppler broadening (a) of hot hydrogen atom and its percentage population (b) corresponding to optical emission observed perpendicular (Position-2) and parallel and perpendicular (Position-2b) to the electric field for pure hydrogen and 50%Ar/ 50%$H_2$ mixture plasma at different radio frequency power levels. (c) The graph shows the normalized $H_\alpha$ emission profile looking perpendicular (Position-2) and parallel (Position-2b) to the electric field for 300 mT 50%Ar/ 50%$H_2$ plasma at 100 W. Note the symmetry of the emission profile.

(a)

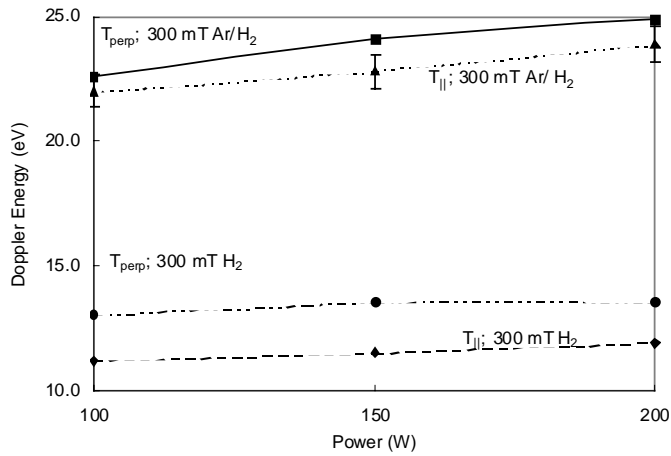

(b)

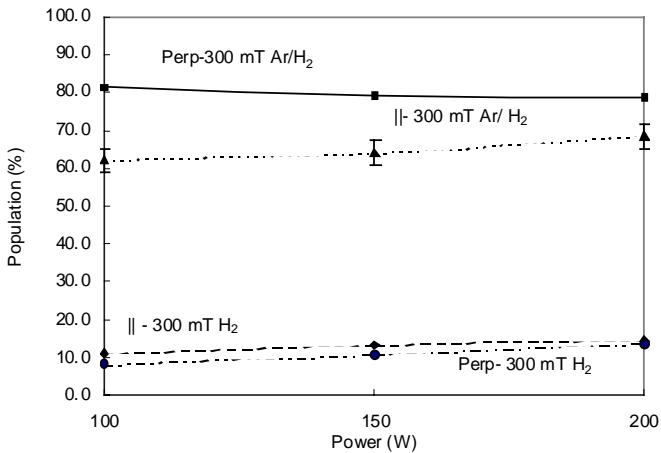

(c)

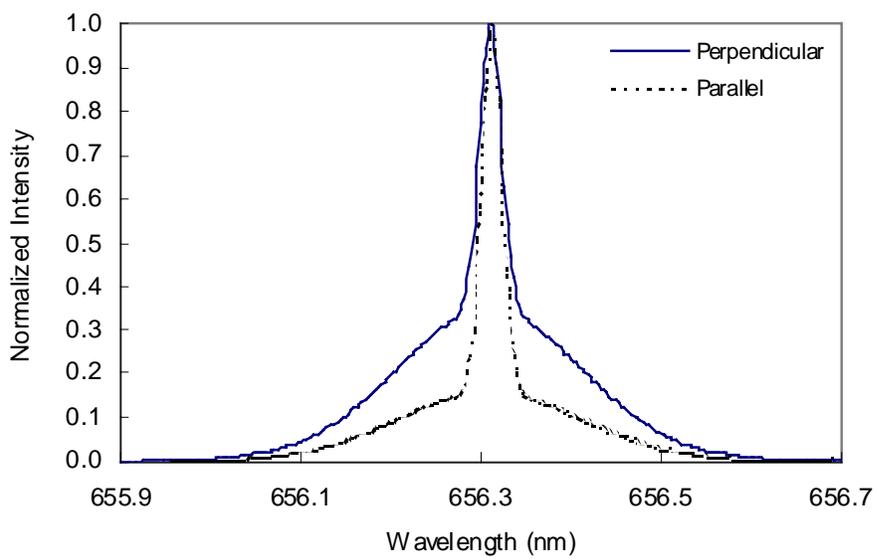

FIG. 10. The temporal behavior of a 17.5% Ar/82.5% H$_2$ plasma was probed. (a) Over time in mixtures containing significant Ar the fraction of atomic hydrogen in the hot state increased to more than 98%. (b) The atomic hydrogen signal strength in this case changed little but the signal from excited argon at 696 nm increased dramatically.

(a)

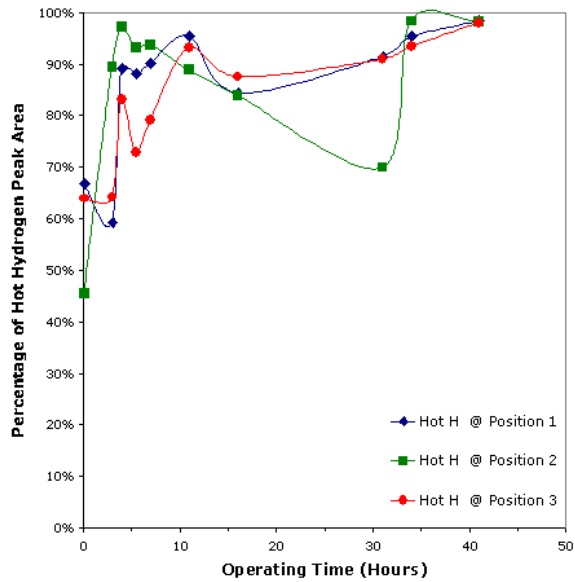

(b)

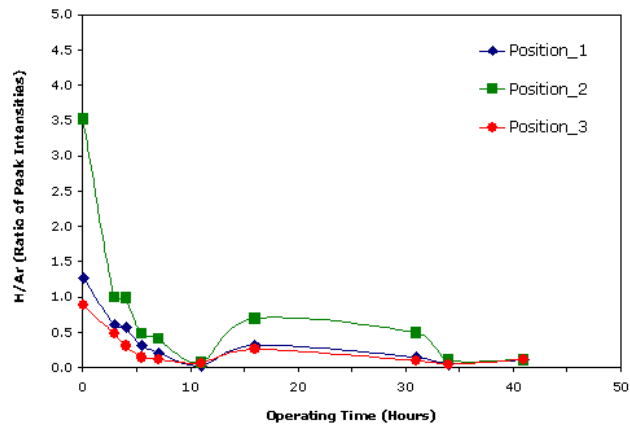

FIG. 11. Temperature probes. All data taken from 100 W 17.5%Ar/82.5% $H_2$, 300 mTorr plasmas, early. (a) The excitation temperature is clearly less than 0.6 eV at all points in the plasma. It is important to note that the excitation temperature can be computed from just the hot fraction, just the cold fraction, or from the total area of the Balmer peaks. There is virtually no difference detected as a function of the approach selected. In the above example, the total peak areas were used. (b) The electron temperature, measured by emission spectroscopy, hence unobtrusively, is clearly less than 0.5 eV everywhere. It is clearly slightly higher at the center of the electrodes, the point with the highest field of the three points measured. The similarity of the magnitudes of the two temperatures (0.3 to 0.6 eV) suggests that all species, other than hot hydrogen atoms, are thermalized in these plasmas, as expected. (c) At all positions within the plasma the temperature of the hot H atoms is far higher than that of the electrons or excitation.

(a)

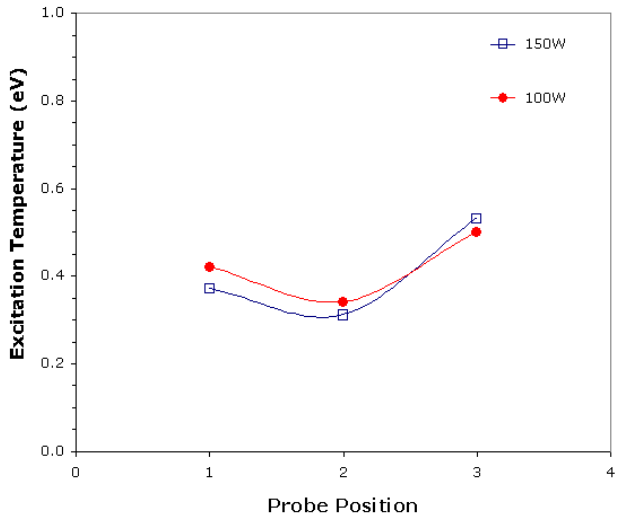

(b)

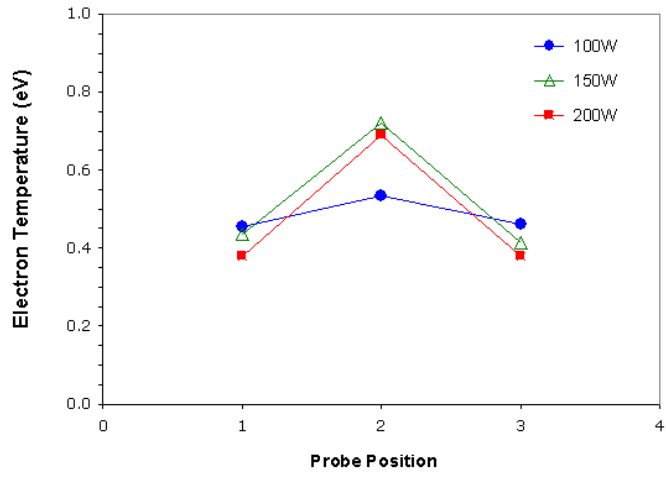

(c)

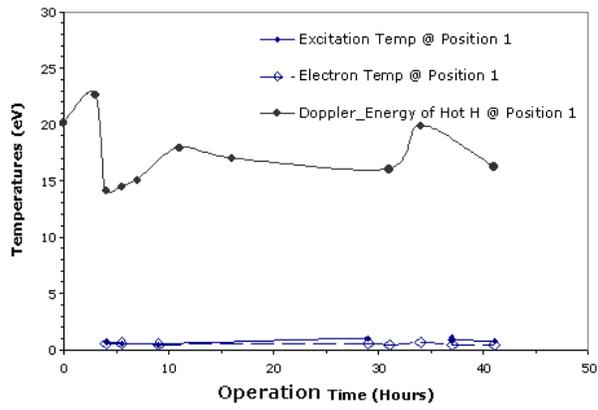

FIG. 12. No significant line broadening was observed in the Balmer $H_\alpha$ line produced in the Xe/$H_2$ (17.5%Xe/82.5% $H_2$) , 100 W, 300 mTorr plasma at any position, either early or late. In order to obtain reasonable signal/noise ratios, counting was performed far longer than for the Ar/$H_2$ plasmas. The 'early' signal shown is about 7% the area of an Ar 17.5%/$H_2$ 82.5% signal collected under the same conditions, early. The 'late' signal is less than 1% the strength of the 'late' Ar 17.5%/$H_2$ 82.5% equivalent. It is clear that the atomic H signal in the control plasma is much smaller than that of the Ar/$H_2$ mixture at all times and that furthermore the H signal in the control plasma drops significantly with time.

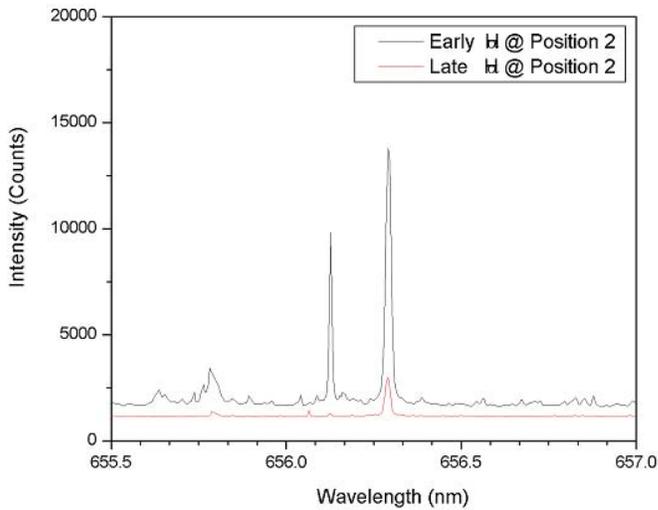